\def\year{2021}\relax
\documentclass[letterpaper]{article} 
\usepackage{aaai21}  
\usepackage{times}  
\usepackage{helvet} 
\usepackage{courier}  
\usepackage[hyphens]{url}  
\usepackage{graphicx} 
\usepackage{graphicx}  
\usepackage{natbib}  
\usepackage{caption} 

\usepackage{bbm}
\usepackage{multirow}
\usepackage{xcolor}
\usepackage{dblfloatfix}
\usepackage{booktabs}
\usepackage{algorithmic}
\usepackage{soul}
\usepackage[linesnumbered,ruled,vlined]{algorithm2e}
\usepackage{eucal}
\usepackage{amsfonts}

\title{\textsc{Phoebe}: Reuse-Aware Online Caching with Reinforcement Learning for Emerging Storage Models}
\author{
 Nan Wu\textsuperscript{\rm 1},
 Pengcheng Li\textsuperscript{\rm 2}\\
}
\affiliations{ \textsuperscript{\rm 1} UC Santa Barbara, Santa Barbara, CA, USA\\ \textsuperscript{\rm 2} Alibaba Group, Sunnyvale, CA, USA \\ Email: nanwu@ucsb.edu, pengcheng.li@alibaba-inc.com
}

\begin{document}
\maketitle

\begin{abstract}

With data durability, high access speed, low power efficiency and byte addressability, NVMe and SSD, which are acknowledged representatives of emerging storage technologies, have been applied broadly in many areas.
However, one key issue with high-performance adoption of these technologies is how to properly define intelligent cache layers such that the performance gap between emerging technologies and main memory can be well bridged.
To this end, we propose \textsc{Phoebe}\footnote{Phoebe is the prophetic Titaness of the Delphic Oracle, which is analogous to our proposed caching policy aiming to approximate the Belady`s optimal algorithm, the oracle of caching policy for fixed-size caches.}, 
a reuse-aware reinforcement learning framework for the optimal online caching that is applicable for a wide range of emerging storage models.
By continuous interacting with the cache environment and the data stream, \textsc{Phoebe} is capable to extract critical temporal data dependency and relative positional information from a single trace, becoming ever smarter over time.
To reduce training overhead during online learning, we utilize periodical training to amortize costs.
\textsc{Phoebe} is evaluated on a set of Microsoft cloud storage workloads.
Experiment results show that \textsc{Phoebe} is able to close the gap of cache miss rate from \texttt{LRU} and a state-of-the-art online learning based cache policy to the Belady's optimal policy by 70.3\% and 52.6\%, respectively.

\end{abstract}

\def \phoebe {\textsc{Phoebe}}

\section{Introduction}
\label{sec:intro}

In the era with ever-growing data expansion, the rapid evolvement in storage technologies, such as the state-of-the-art non-volatile memory express (NVMe) and the solid-state drive (SSD), offers strong support to the high-performance, low-power and large-capacity data storage tiers in data centers.
Despite tremendous endeavor made, the storage I/O is still the performance bottleneck due to the large disparity in latency between fast host-side memories and slow back-end storage drives, which causes incredibly long I/O waiting time and a substantial amount of CPU idle wastage.
This situation is even exacerbated by the fact that many cloud applications are I/O intensive, such as video streaming, news feeds and social networks~\cite{Hoseinzadeh:corr19}.


To alleviate this disparity in latency, caching tiers have been widely employed to reside between main memories and back-end storage drives.
The performance of caching systems is usually influenced by three factors: the data allocation policy, the accuracy of data hotness recognition and the data eviction policy.
The data allocation policy basically controls the data flow and determines admissions of various data, such as read-only, write-only, or both;
the high accuracy of data hotness recognition can prevent cache pollution from unnecessary data, improving cache performance via locality protection;
the data eviction policy decides which data block to evict when the cache is full, thus indirectly increasing the effective cache capacity.
These three factors, which give attention to three different aspects, are highly relevant.
One thing worth mentioning is that conventional caching policies, such as \texttt{LRU} and \texttt{LFU}, are not a one-for-all solution complying with all these factors \cite{Li+:ASPLOS19,LiG:arxiv20,liu2020imitation}.


With these considerations, there arise increasing interests seeking for a unified workaround of optimized caching.
In the meantime, both the variety and size of modern workloads are drastically growing, which urges the cache systems to equip with learning capabilities such that they can adjust their behaviors in real time according to versatile working patterns.
The recent advancement in reinforcement learning (RL) has paved the way for solving many complicated problems, including but not limited to robotics \cite{lillicrap2015continuous,gu2017deep,hwangbo2019learning}, video game playing \cite{jaderberg2019human,vinyals2019grandmaster}, autonomous driving \cite{sallab2017deep,kiran2020deep} and neural architecture search \cite{baker2016designing,tan2019efficientnet}. 
Even though there has been a series of investigations applying machine learning for system optimizations \cite{Dean:NIPS17,Lecuyer+:HOTNETS17,Bychkovsky+:MS2018,Song+:NSDI20}, there is an absence of RL caching systems competitive with existing heuristics, due to two principal unresolved issues: the lack of adequate Markov property in the problem formulation and the prohibitively large training overhead during the online learning process.
First, in order to achieve better learning efficiency, the problem formulation (especially the state formulation) should obtain great Markov property.
This indicates that the state representation should involve enough information extracted from proper features for the RL agent to make the optimal decisions.
Nevertheless, under the cache scenario, increasing information floods into the observation as the data access trace proceeds, making it difficult to distill useful information for an appropriate state representation.
Simultaneously, the state representation has to be compatible with the invariable deep neural network (DNN) input/output structure, which has proved to be powerful function approximators in RL.
Second, training overhead is always a concern for online learning based caching systems.


To this end, we propose a reuse-aware RL framework for optimized online caching, namely \phoebe, leveraging the deep deterministic policy gradient (DDPG) \cite{silver2014deterministic,lillicrap2015continuous} to assign priorities to each data block such that the overall hit rate is maximized.
\phoebe, which exploits reuse-based data locality that is intrinsic to data access patterns, is equipped with carefully crafted neural networks as function approximators, and by continuous interacting with the cache environment and the data stream, it is capable to extract critical temporal data dependency and relative positional information \emph{from a single trace}, making itself ever smarter over time.
To strike a balance between the Markov property and cost of state representation, we consider the history information from a past fixed-length sequence of accesses preceding to the up-to-date data access.
To make our design online, \phoebe\ also adopts periodical training to amortize run-time overhead. 


In summary, this paper makes the following contributions:
\begin{itemize}
\item We propose \phoebe\, a reuse-aware RL framework for the optimal online caching through DDPG.
\item We present a rather fast cache interface designed with $O(1)$ time complexity, enabling high-efficiency cache environment simulations.
\item \phoebe\ is able to distill adequate information simply \textit{from a single cache trace}, which would be universally applicable in theory and can be generalized for different cache scenarios.
\item In evaluation over a set of real-world Microsoft cloud workloads, \phoebe\ is able to close the gap of the cache miss rate from \texttt{LRU} and a state-of-the-art online learning based cache policy to the Belady's optimal policy by 70.3\% and 52.6\%, respectively.
\end{itemize}


\section{Preliminaries}
\label{sec:pre}

\subsection{Storage caching tiers}
\label{sec:back-cache}

\begin{figure}[t]
\centering
\includegraphics[width=\linewidth]{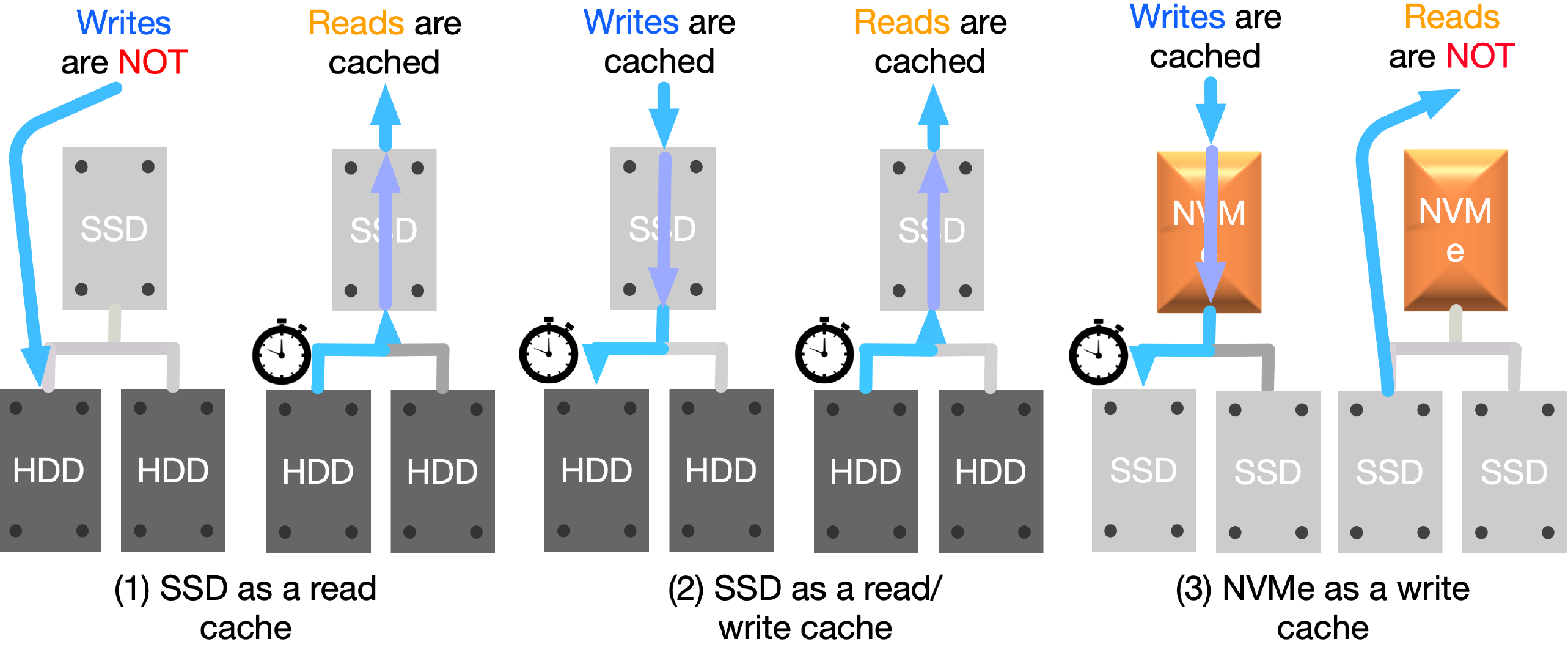}
\caption{Different caching scenarios with SSD and NVMe.} 
\label{fig:ssd-nvm-cache}
\end{figure}

Storage drive technologies have been evolving greatly over years.
HDD devices are currently surpassed by the faster and more reliable SSD devices, and emerging persistent memory devices that adopt NVMe are believed to have great potentials in the future market \cite{jung2020openexpress}.
In latency-wise, the NVMe is hundreds of times faster than SSD devices, and SSD devices are hundreds or even thousands of times faster than HDD devices.
In price-wise, HDD devices have much lower prices and the NVMe is the most expensive one in terms of price per byte.
Regarding different features of these storage devices, HDD devices are often deployed as storage-level capacity drives in data centers taking advantage of their extremely large capacity (i.e., up to tens of tera-bytes capacity), while the NVMe and SSD devices are employed as cache drives due to their fast speed and relatively high cost.
The special characteristics of the NVMe and SSD devices make them possible to accommodate various requirements of cache designs in different storage hierarchies.
Figure \ref{fig:ssd-nvm-cache} shows three common practices applying the NVMe and SSD devices as caching tiers in data centers.

\textbf{SSD as a read/write cache for HDD.}
It is a common case to employ SSD devices as the read/write cache to accelerate read and write operations in HDDs by hundreds of folds \cite{huang2016improving}.
The read cache stores recently and frequently read data for faster access, minimizing random traffics to HDD devices that usually incur significant latency caused by seeking and rotational delays.
The write cache can absorb bursts of writes and coalesce writes/re-writes to minimize the cumulative traffic to capacity drives.
SSD caches often consider the write-back policy due to  data persistence.



\textbf{SSD as a read-only cache for HDD.} 
Such kind of cache architecture aims to enhance the lifespan of SSD devices by redirecting the writing stream to fetching data directly from HDD devices.
In this case, the cache capacity can be completely utilized for read operations that are on the critical paths of application executions, significantly reducing the latency.
These read-only caches are mostly specifically optimized for read-critical workloads, such as iPhone store downloads and news feeds.


\textbf{NVMe as a write-only cache for SSD.}
SSD devices have been widely deployed in large-scale enterprise storage or data centers \cite{do2019programmable,maneas2020study}.
In terms of caching for SSD devices, the NVMe is often recommended as a write-only cache between the main memory and SSD devices \cite{fan2014h,lee2018empirical}, since NVMe is much faster and can endure orders of magnitude more writes than that of SSD devices.
Thus, the lifespan of SSD devices can be significantly extended, as a number of writes can be coalesced in the cache and then drained as needed.
This type of caches are usually specifically dedicated for write-sensitive workloads, such as tweets and photo sharing.


\subsection{Reinforcement Learning}
\begin{figure}[tbp]
  \centering
  \includegraphics[width=0.85\linewidth]{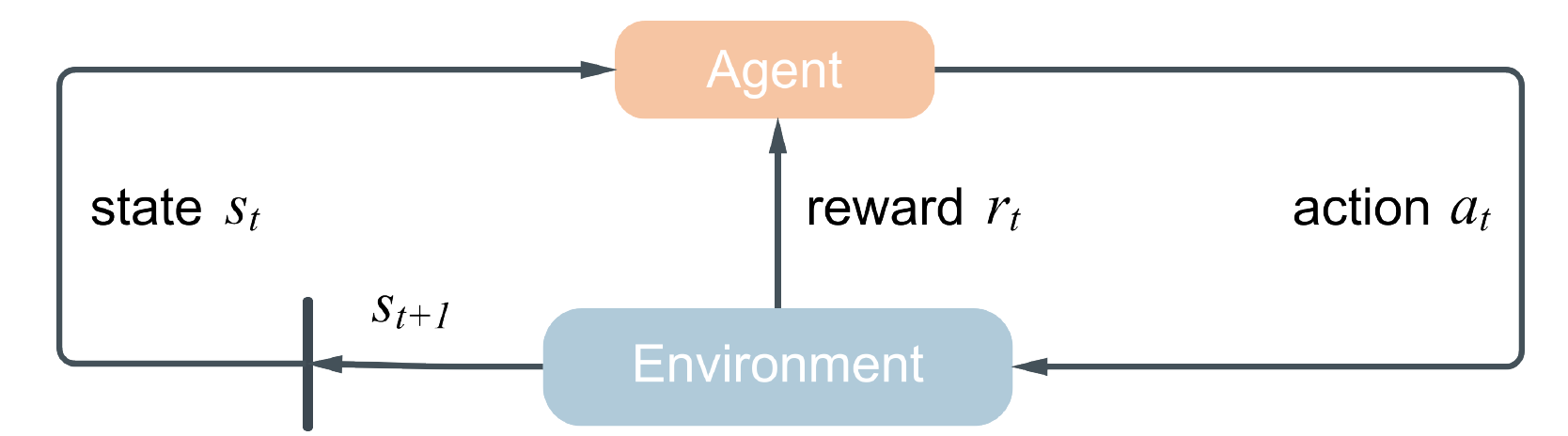}
  \caption{A typical framing of RL.}
  \label{fig:rl0}
\end{figure}

In the standard setting of RL \cite{sutton2018reinforcement}, an agent interacts with an environment $\mathcal{E}$ over a number of discrete time steps, as shown in 
Figure \ref{fig:rl0}.
At each time step $t$, the agent observes a state $s_t$ from the \textit{state space} $\mathcal{S}$, and selects an action $a_t$ from the \textit{action space} $\mathcal{A}$ according to its policy $\pi$, which is a mapping from the state $s_t$ to the action $a_t$. 
In return, the agent receives the next state $s_{t+1}$ and a scalar reward $r_t: \mathcal{S}\times\mathcal{A}\to\mathbb{R}$. 
This process continues until the agent reaches a final state after which the process restarts.
The accumulated rewards after the time step $t$ can be formulated as $R_t={\sum_{k=0}^\infty \gamma^kr_{t+k}}$, where $\gamma \in(0,1]$ is the \textit{discount factor}.
The state-action value $Q_{\pi}(s,a)=\mathbb{E_\pi}\lbrack R_t|s_t=s, a_t=a\rbrack$ is the expected return after selecting action $a$ at state $s$ with policy $\pi$.
Similarly, the state value $V_{\pi}(s)=\mathbb{E_\pi}\lbrack R_t|s_t=s\rbrack$ is the expected return starting from state $s$ by following policy $\pi$. 
The goal of the agent is to maximize the expected return for every state.

The approaches to RL can be categorized as value-based methods or policy-based methods.
In the value-based methods, the state-action value function $Q_{\pi}(s,a)$ is approximated by either tabular approaches or function approximations, and at each state $s$ the agent always selects the optimal action $a^*$ that can bring the maximal state-action value $Q_{\pi}(s,a^*)$, with one well-known example as Q-learning \cite{watkins1992q}.
Among those policy-based methods, policy gradients have been broadly applied under different RL scenarios, where the basic idea is to directly parameterize the policy via a probability distribution $\pi_{\theta}(s,a)=\mathbb{P}(a|s;\theta)$ that stochastically selects the action $a$ given the state $s$ according to the parameters $\theta$.

In order to better handle continuous action spaces, there presents the deterministic policy gradient (DPG) \cite{silver2014deterministic,lillicrap2015continuous} that aims to learn a deterministic policy $\mu_{\theta}(s):\mathcal{S}\rightarrow\mathcal{A}$ by maximizing $\mathcal{J}(\mu_{\theta})=\mathbb{E}\left[ \sum_{k=0}^\infty \gamma^k r_{k}|\mu_{\theta}\right]$, where $0<\gamma<1$.
Then the deterministic policy gradient can be derived as: 
\begin{equation}
    \label{eq:gradient}
    \nabla_{\theta}\mathcal{J}(\mu_{\theta})=
    \mathbb{E}\left[ \nabla_{\theta}\mu_{\theta}(s) \nabla_a Q_{\mu}(s,a)|_{a=\mu_{\theta}(s)}\right].
\end{equation}

DPG is often combined with the actor-critic algorithm, in which the \textit{critic} estimates the state-action value function $Q_w(s,a)\approx Q_{\mu}(s,a)$ by adjusting parameters $w$ based on Q-learning and the \textit{actor} learns the deterministic policy $\mu_{\theta}(s)$ by ascending the gradient of the state-action value function.

\section{Approach}
\subsection{Formulation of Reuse-Aware Caching}
For a given data access trace, the reuse distance of a data block is defined as \emph{the number of accesses between the current access and the consecutive next access to the same data block}, measuring the data access locality.
Taking advantage of the reuse distance, the Belady`s optimal algorithm \cite{Belady66}, which always discards the data block that will be accessed the furthest in the future, i.e., evicting the data block having the longest reuse distance at that timestamp, is the oracle of the maximal hit rates for fixed-size caches. However, the Belady`s optimal algorithm is infeasible in real since the future can never be known in advance. 

In this work, rather than explicitly predicting reuse distances, which could be unbounded values and thus increase the difficulty and complexity for high-accuracy predictions, we predict a newly defined indicator, \textit{stay priority}, to represent the relative importance of each data block.
Then the cache replaces data blocks according to their stay priority values when eviction events happen (i.e., evicting the data block with the lowest stay priority), aiming to maximize the cache hit rate. 
The stay priority values have time lagging property, i.e., high priority values at stale timestamps cannot stay as high at up-to-date timestamps, and thus they are often combined with corresponding timestamps to derive the final eviction decision.

\begin{figure}[tbp]
  \centering
  \includegraphics[width=\linewidth]{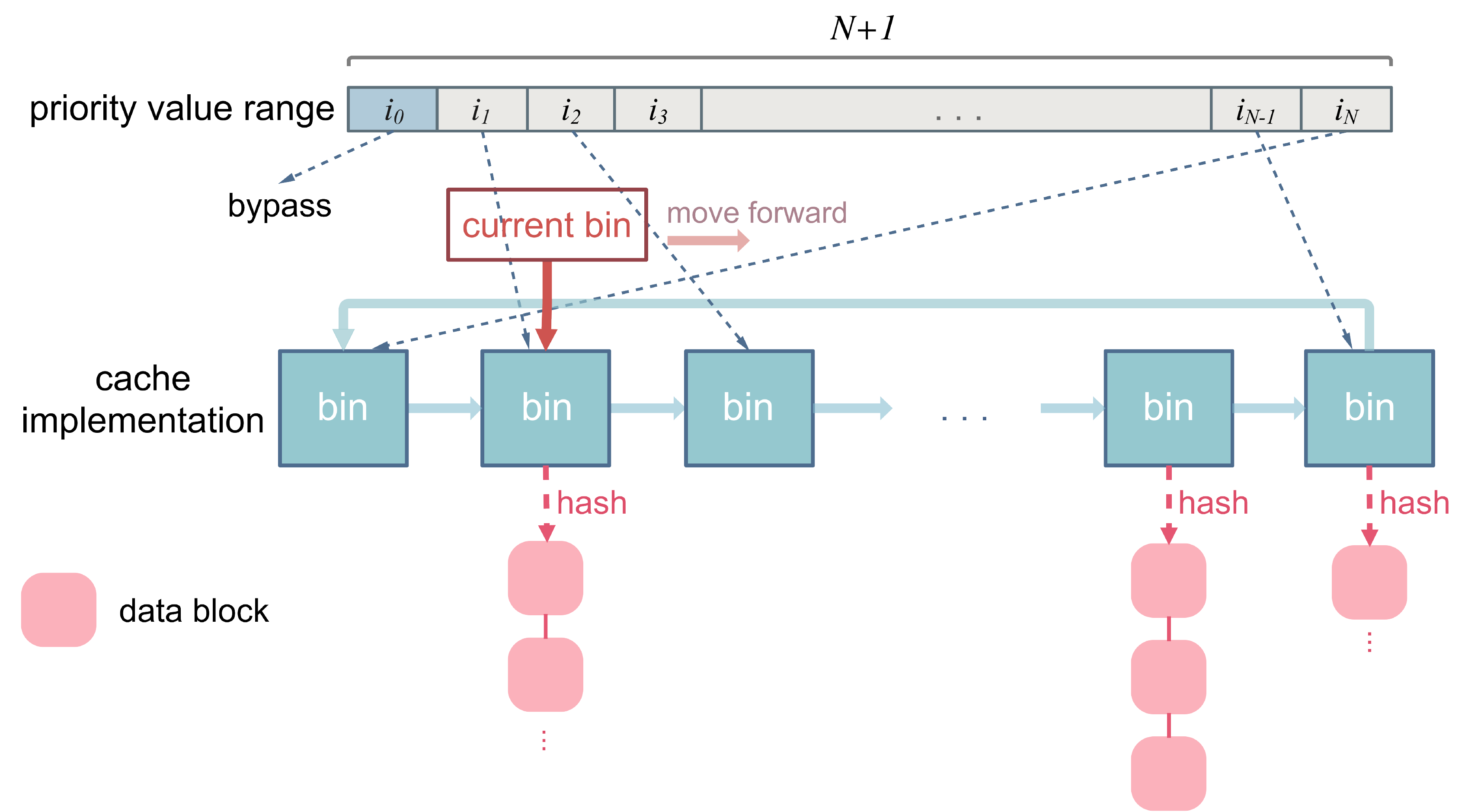}
  \caption{Cache environment implementation.}
  \label{fig:cache_bins}
\end{figure}

With the demand on a fast cache interface, we design the cache as a set of $N$ bins concatenated in a circular manner, where data blocks in the same bin obtain indiscriminate priorities.
There is a pointer pointing to the currently first bin (denoted as $b_{first}$), which is maintained and updated regarding the moving current timestamp (see Figure \ref{fig:cache_bins}).
As the trace proceeds, a stay priority value bounded in a certain range will be assigned for each accessed data block. 
Assume that the output range of priority values is $[p_{low}, p_{up}]$, and this range is evenly partitioned into $N+1$ intervals denoted with indices as $i_0,i_1,i_2, ..., i_N$.
When there happens a cache miss, if the priority value of the currently referenced data block falls in the interval $i_0$, this data block will be bypassed and not get access to the cache if the cache is full;
in other cases, if the priority value falls in the interval $i_k$ ($1\leq k\leq N$), then this data block will be placed in the bin indexed with $(b_{first}+k-1) \bmod N$.
As for a cache hit, the position of the referenced data block will be updated in a similar way.
In case of evictions, the data blocks in the bin $b_{first}$ will be discarded based on their order of entrance, and once this bin is empty, the pointer moves to the next non-empty bin.
One thing worth noting is that if the priority values for all data blocks are the same constant, this caching scheme degrades to \texttt{LRU}, guaranteeing the performance no worse than \texttt{LRU}.

In order to correctly predict priority values, we focus on 9 features to extract reuse information that considers both global and local patterns.
The first six features and the last one are global features; the $7^{th}$ and $8^{th}$ features are local features that are collected from a sliding window preceding to the current data block.
\begin{itemize}
    \item \textbf{Data block address.} The first feature is the data address or data block ID. For example, in the CPU cache scenario, it is the memory address.
    \item \textbf{Data block address delta.} This feature is computed as the difference between the current data block address and the previous one, with the first address delta set as zero.
    \item \textbf{Frequency.} The access frequency of the current data block is counted over the entire history of the trace.
    \item \textbf{Reuse distance.} The reuse distance illustrates the recency information of data blocks, which is computed by a hash table recording the last access time of each data block.
    \item \textbf{Penultimate reuse distance.} The second last reuse distance is supplementary to reuse distance for better leverage of data recency.
    \item \textbf{Average reuse distance.} This feature is the expected reuse distance over the entire history, and can be recognized as an estimate of the next potential reuse distance.
    \item \textbf{Frequency in the sliding window.} The access frequency of the current data block over the past $H$ accesses preceding to itself is counted.
    \item \textbf{The number of cache misses in the sliding window.} We count the number of cache misses of the current data block over the past $H$ accesses preceding to itself. The more cache misses for one data block, the more compensation we need to make in its priority value.
    \item \textbf{Priority value.} We consider the priority value of the current data block. Once this value is available, it is referenced to make future predictions.
\end{itemize}

\subsection{Online Caching with DDPG}
For online caching, the agent attempts to learn an optimal policy of assigning stay priorities to data blocks such that the overall hit rate is maximized;
the cache environment gives feedback to the agent by different rewards (i.e., cache hits and misses) to encourage or punish the agent according to its behaviors.
By continuous interactions with the cache environment, the agent is able to learn the optimal policy.

\begin{figure}[tbp]
  \centering
  \includegraphics[width=\linewidth]{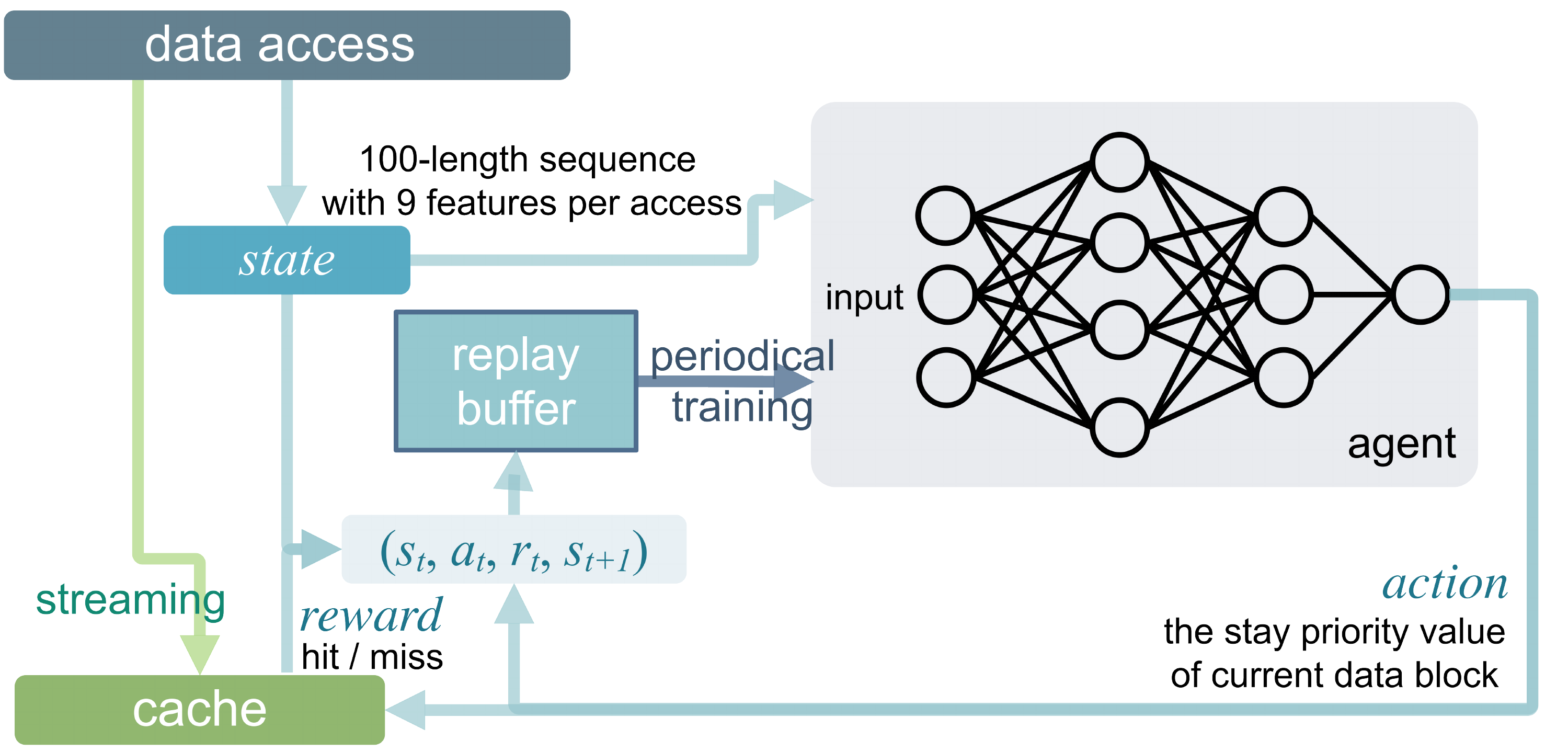}
  \caption{Overview of online caching optimization with RL.}
  \label{fig:framework}
\end{figure}

We model the online caching problem as a \emph{Markov decision process}.
At each time step $t$, the agent observes the history information as of the current access, which is referred to as the \textit{state} $s_t$ in the state space $\mathcal{S}$. 
It is undoubtedly that taking the entire history into consideration would be most beneficial for the agent to make decisions from the perspective of Markov property, but this is prohibitively expensive to represent the states as more and more accesses accumulate.
Aiming to find a trade-off between the Markov property and the cost of state representations, we consider the history information over a fixed-length sequence of accesses preceding to the most recent data access.
Given the state $s_t$, a stay priority value for the being accessed data block is generated, which is referred to as the \textit{action} $a_t$ in the action space $\mathcal{A}$.
With this value, the cache environment is able to decide whether to give admission to this data block and which data block to evict if the cache is full.
The cache hit or miss in the next time step is regarded as the immediate reward to tune behaviors of the agent.
These rewards, together with information from states, actions, and state-action values, are combined to train the agent and update its following policy.
The overview is shown in Figure \ref{fig:framework}.

In a nutshell, the mathematical representations of the state, action and reward of the online caching problem are detailed as follows.
\begin{itemize}
    \item \textbf{State.} The state $s_t$ encodes the locality information from a past fixed-length sequence of accesses just preceding to the current data block. We empirically set the considered sequence length as $100$. Since each access has $9$ features as aforementioned, the $100$-length sequence composes a $9 \times 100$ matrix as a state.
    While observing the current data block, its priority value is not yet generated and we manually place it as zero in the state representation.
    \item \textbf{Action.} The action $a_t$ is to assign a stay priority value ranging within $[-1,1]$ to the current data block, which will be later used by the cache to make eviction decisions.
    \item \textbf{Reward.} The reward gives immediate feedback from the environment to the agent, indicating whether the next referenced data block is in the cache. A cache hit is rewarded with $+1$ while a cache miss is penalized with $-1$.
\end{itemize}

\begin{figure}[tbp]
  \centering
  \includegraphics[width=\linewidth]{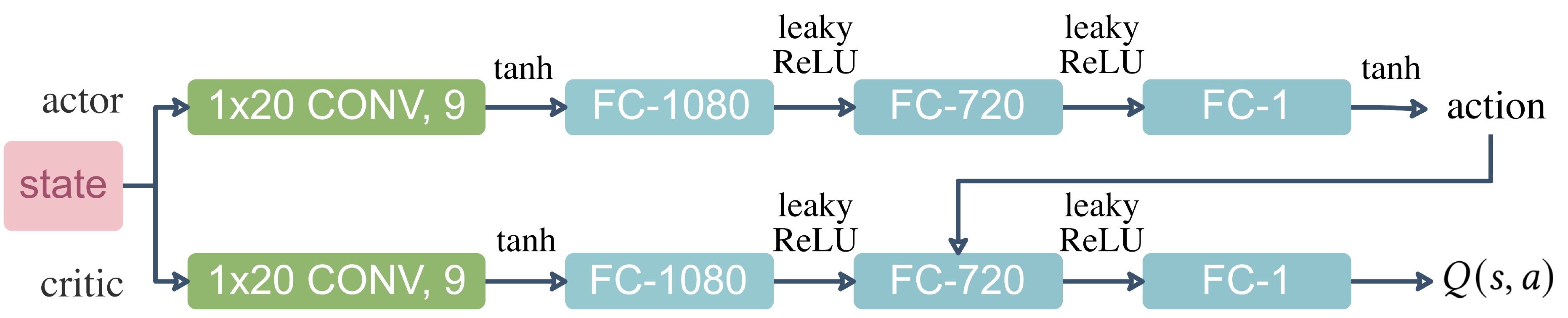}
  \vspace{-18pt}
  \caption{DNN structure of the RL agent: the actor and the critic.}
  \vspace{-8pt}
  \label{fig:dnn}
\end{figure}

In our work, we apply the DDPG with the actor-critic algorithm.
As shown in Figure \ref{fig:dnn}, both the actor and critic have similar network structures.
The input to these DNNs is the current state, and then the actor outputs the priority value of the current data block in scalar.
Since the state-action value is a function of both the current state and the action being taken, the output of the actor is merged to the critic after its first fully-connected (FC) layer.
We employ multiple $1 \times 20$ convolutional filters to extract temporal features from history information, since each row of our state representation (i.e., 9 $\times$ 100 matrix) denotes how one of the $9$ considered features behaves during the past $100$ accesses.
The batch normalization is applied after each layer to avoid over-fitting and reduce sensitivity of initialization; 
the activation function is $tanh$ for CONV layers and leaky ReLU with $\alpha=0.1$ for FC layers;
the output of the actor also uses $tanh$ to bound priority values in $[-1,1]$.
The network parameters are learned by Adam optimizer.
To leverage past experiences, we utilize the experience replay with a replay buffer storing tuples $(s_t,a_t,r_t,s_{t+1})$ from history trajectories; 
to sufficiently explore the large search space, we add Ornstein-Uhlenbeck noise \cite{uhlenbeck1930theory} to the action space.

It is noticeable that to reduce overhead during online learning, we leverage \textbf{\textit{periodical training}} to amortize cost, i.e., training on 5 consecutive accesses per 100 accesses.
The entire procedure is summarized in Algorithm \ref{alg:rl}.

\begin{algorithm}[t]
\SetAlgoLined
\LinesNumbered
Initialize the actor $\mu_{\theta}(s)$ and the critic $Q_w(s,a)$ with weights $\theta$ and $w$ \;
Initialize the target networks $\mu_{\theta}'$ and $Q_w'$ with weights $\theta' \gets \theta$ and $w' \gets w$\;
Initialize an Ornstein-Uhlenbeck process $\mathcal{N}$ for action exploration\;
Initialize the timestamp $t \gets 0$ and receive initial state $s_t$ after observation of the first data block\;
\While{trace not end}{
  Generate the priority value for current data block by action $a_t=\mu_{\theta}(s_t)+\mathcal{N}_t$ with current policy and noise\;
  Observe the next referenced data block, i.e. the $(t+1)^{th}$ data block, to receive the new state $s_{t+1}$ and reward $r_t$\;
  Store $(s_t,a_t,r_t,s_{t+1})$ into replay buffer $\mathcal{R}$\;
  \If{$ t \bmod 100 \in [95,99]$}{
  Sample a minibatch of $\mathcal{K}$ transitions from $\mathcal{R}$\;
  Set $y_i=r_i+\gamma Q_w'(s_{i+1},\mu_{\theta}'(s_{i+1}))$\;
  Update the critic by minimizing the loss: $L=\frac{1}{\mathcal{K}} \sum_{i} (y_i - Q_w (s_i,a_i))$\;
  Update the actor policy by $\nabla_{\theta}\mathcal{J}\approx\frac{1}{\mathcal{K}} \sum_{i} \nabla_a Q_w(s,a)|_{s=s_i,a=\mu_{\theta}(s_i)} \nabla_{\theta}\mu_{\theta}(s_i)$\;
  Update the target networks:
  $\theta' \gets \tau \theta +(1-\tau)\theta'$,
  $w' \gets \tau w+(1-\tau)w'$\;
  }
  $t \gets t+1$\;
 }
 \caption{Deep deterministic policy gradient for online caching optimization.}
 \label{alg:rl}
\end{algorithm}

\section{Experiment}
\subsection{Experiment Setup}
\paragraph{Workloads.}
We evaluate \phoebe\ over a set of storage traces, which collect active disk block accesses and are profiled from six different production servers in the Microsoft cloud \cite{Narayanan+:TOS08}, as summarized in Table \ref{table1}.

\begin{figure*}[htbp]
  \centering
  \vspace{-10pt}
  \includegraphics[width=0.95\linewidth]{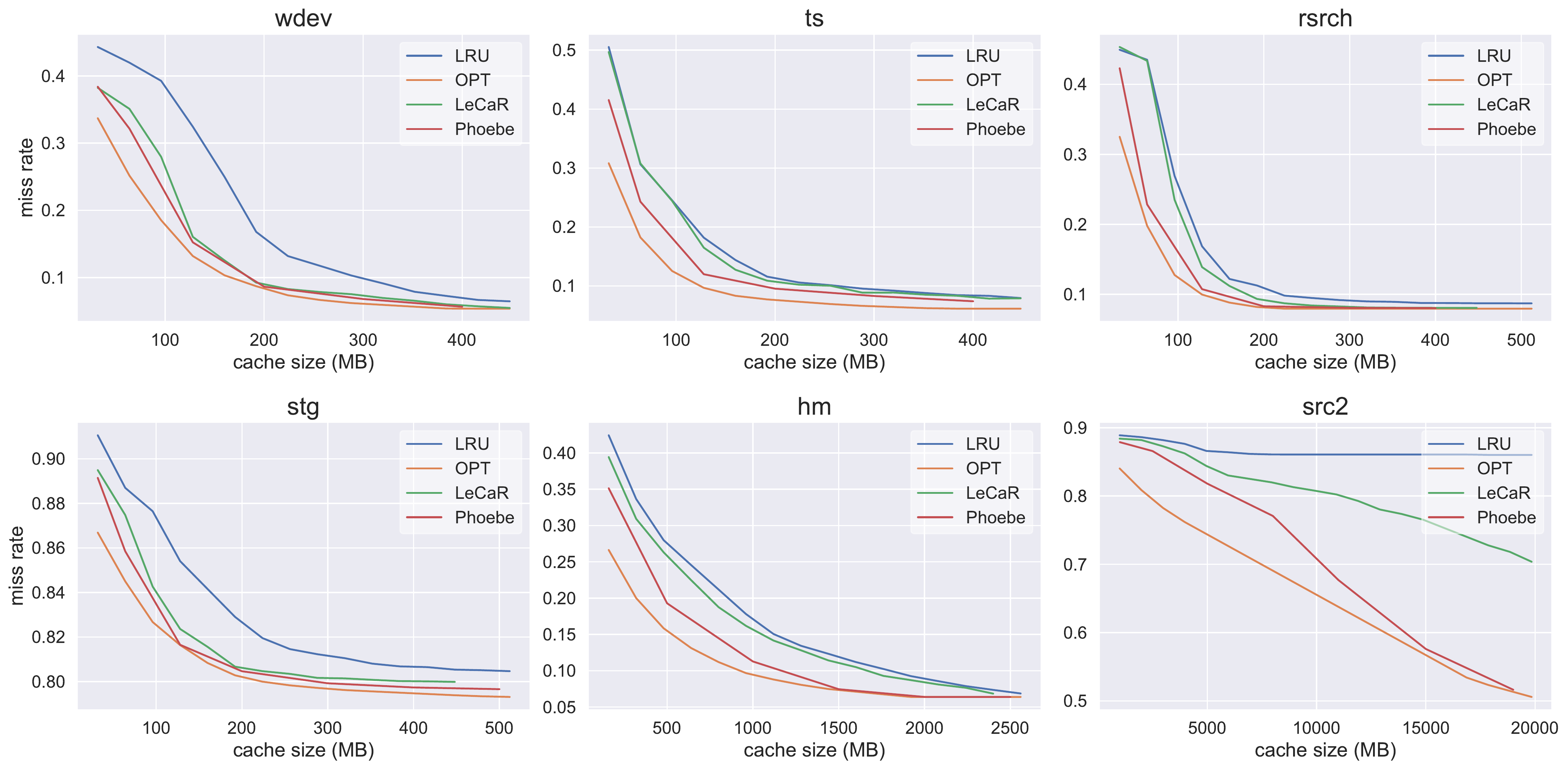}
  \vspace{-10pt}
  \caption{Performance comparison for 6 MSR traces.}
  \vspace{-5pt}
  \label{fig:result}
\end{figure*}

\begin{figure*}[!h]
  \centering
  \includegraphics[width=0.95\linewidth]{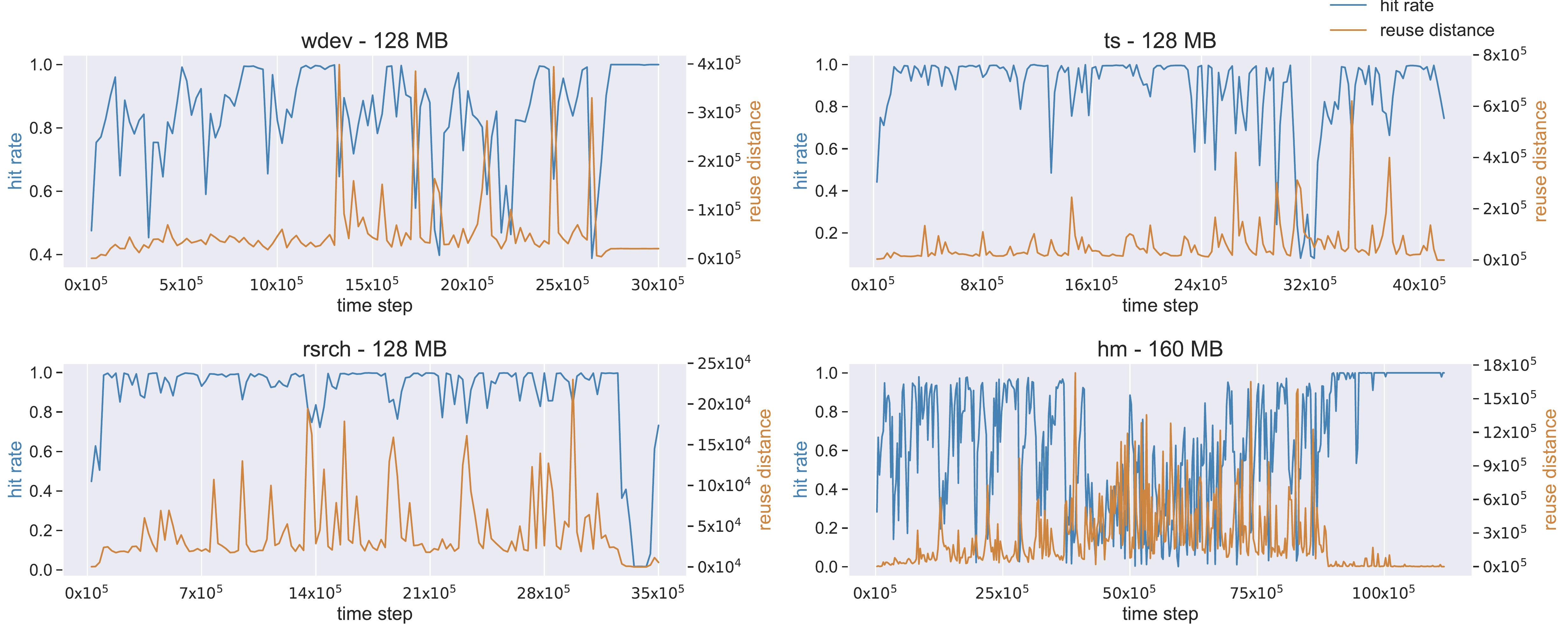}
  \vspace{-8pt}
  \caption{Hit rate and reuse distance as traces proceed (smoothed by averaging on every 25,000 steps with cache size marked).}
  \vspace{-8pt}
  \label{fig:hit_reuse}
\end{figure*}

\begin{table}[]
\centering
\small
\begin{tabular}{|c|c|c|c|}
\hline
\textbf{Domain  (\# volumes)} & \textbf{Name} & \textbf{Trace length} & \textbf{\#Data blocks} \\ \hline
test web server (4)     & wdev  & 3,024,140  & 162,629    \\ \hline
terminal server (1)     & ts    & 4,181,323  & 256,922    \\ \hline
research projects (3)   & rsrch & 3,508,103  & 279,128    \\ \hline
hardware monitoring (2) & hm    & 11,183,061 & 715,049    \\ \hline
source control (3)      & src2  & 28,997,811 & 10,939,638 \\ \hline
web staging (2)         & stg   & 28,538,432 & 22,608,572 \\ \hline
\end{tabular}
\caption{Trace characteristics. \textit{Trace length} measures the number of accesses in a trace, and \textit{\#Data blocks} denotes the number of unique data blocks accessed. The data block size is set as 4KB for all traces.}
\vspace{-10pt}
\label{table1}
\end{table}

\vspace{-10pt}
\paragraph{Baselines.}
\begin{itemize}
    \item \texttt{LRU} (least recently used): a classical cache replacement policy that always evicts the least recently used data blocks, capturing data recency.
    \item \texttt{OPT} \cite{Belady66}: the Belady`s algorithm, the optimal caching policy for fixed-size caches, which provides the theoretical upper bound of hit rates.
    \item \texttt{LeCaR} \cite{vietri2018driving}: an online learning based caching policy that switches between \texttt{LRU} and \texttt{LFU} (least frequently used) according to learned probabilities.
    \item \textsc{Parrot} \cite{liu2020imitation}: an offline imitation learning based caching policy that mostly focuses on hardware caches. It requires to run \texttt{OPT} on a predefined trace in advance so that it can imitate behaviors of \texttt{OPT}, and thus it is not an online approach.
\end{itemize}

\vspace{-15pt}
\paragraph{Configurations.}
In \texttt{LeCaR}, we set the learning rate as $0.45$, with the discount rate as $0.002^{1/\sqrt[4]{C}}$, where $C$ is the cache size in the number of data blocks.
Since \textsc{Parrot} is an offline policy designed for CPU caches, which also takes \textit{program counters} ($pc$) as an input feature and considers hardware characteristics such as the cache associativity, we conduct simple comparison as demonstration.
The cache is considered full-associative with the size as 4MB (and data block size as 4KB); the $pc$ for each data block is manually set as zero since software caches do not have this feature.
We consider the first 20,000 accesses intercepted from 4 MSR traces, and conduct experiments on these four partial traces.

As for hyperparameters in \phoebe, the learning rates of the actor and critic are set as $\alpha_{\theta}=0.02$ and $\alpha_w=0.005$, with the soft update factor $\tau=0.002$.
The size of minibatch is $\mathcal{K}=64$.
We test multiple discount factors, i.e., $\gamma \in [0.925,0.950,0.970,0.990]$, since the discount factor, which depicts the relative weights of immediate rewards and long-term expectations, has conspicuous influence on learning performance under different cache sizes.
The sliding window width is also set as $H=100$.

\subsection{Results and Discussions}
In our evaluation, we give attention to two major metrics, the absolute cache hit/miss rate that is closely correlated to system performance, and the relative cache miss rate that depicts how much of the gap between \texttt{LRU} or \texttt{LeCaR} and \texttt{OPT} is mitigated.
Given a policy with a miss rate $r$, the relative miss rates to \texttt{LRU} and \texttt{LeCaR} are defined as $\frac{r_{lru}-r}{r_{lru}-r_{opt}}$ and $\frac{r_{lecar}-r}{r_{lecar}-r_{opt}}$, where $r_{lru}$, $r_{lecar}$, and $r_{opt}$ are miss rates of \texttt{LRU}, \texttt{LeCaR} and \texttt{OPT}, respectively.

Figure \ref{fig:result} compares the cache performance of four cache replacement policies on 6 MSR traces.
Among all the traces, \phoebe\ achieves significantly lower cache miss rates than \texttt{LRU}, and clearly outperforms \texttt{LeCaR}.
Generally, to achieve comparable levels of cache miss rates as \phoebe\ by using \texttt{LRU} would require increasing the cache capacity to $2-3 \times$.
In the case of $src2$, \texttt{LRU} is stuck at a high miss rate, around $86\%$, even with large cache sizes, whereas \phoebe\ gets a sharp drop after the cache size reaches 2.5GB.
As for the relative cache miss rate averaged over all traces across various cache sizes, \phoebe\ is able to close the gaps from \texttt{LRU} and \texttt{LeCaR} to \texttt{OPT} by 70.3\% and 52.6\%, respectively.

Through continuous interactions with the cache environment and varying workloads at runtime, \phoebe\ adjusts its policy to approximate the optimal caching policy.
Figure \ref{fig:hit_reuse} illustrates how the local hit rate changes with different reuse distance situations over time.
Since the reuse distance provides information of locality patterns intrinsic to data access traces to some extent, these curves show how \phoebe\ adapts to dynamic characteristics of workloads.
With stable or mildly changing reuse distance curves, the hit rate curves maintain a relatively high level (e.g., the first half of $wdev$ and $ts$);
once there arise spikes in reuse distance curves, indicating sudden changes of data access patterns, the corresponding hit rate curves drop sharply and then they gradually recover to high-hit-rate status as \phoebe\ adapts its cache policy to different patterns.
It is noteworthy that \phoebe\ has strong online learning capability regarding ever-changing workloads, with one evidence from the middle part on $hm$ that even though the reuse distance fluctuates drastically there is still an increasing trend of the local hit rate.

\begin{figure*}[htbp]
  \centering
  \includegraphics[width=.98\linewidth]{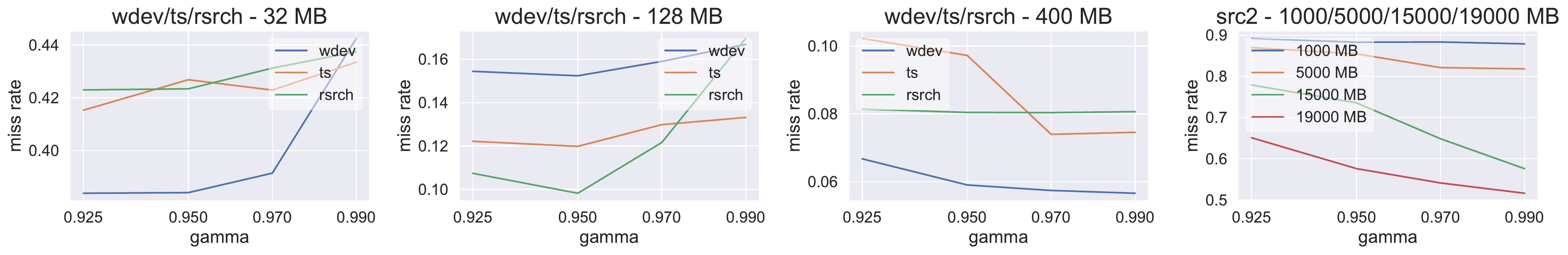}
  \vspace{-8pt}
  \caption{Influence of different discount factors $\gamma$ on cache miss rates, under different cache sizes.}
  \label{fig:gamma}
\end{figure*}

\begin{figure}[htbp]
  \centering
  \includegraphics[width=0.95\linewidth]{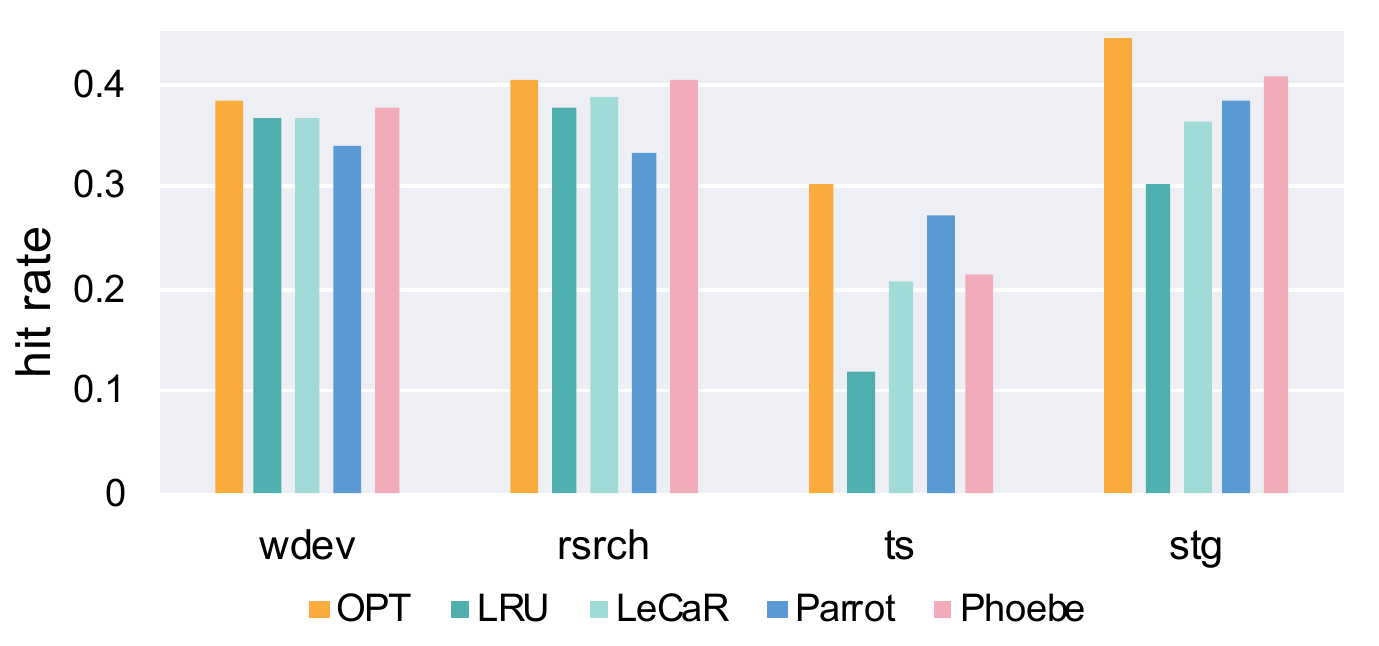}
  \caption{Comparison over 4 MSR traces on the first 20,000 accesses, with the cache size as 4MB.}
  \label{fig:parrot}
\end{figure}

As aforementioned, we test multiple discount factors to see their influence on learning performance.
Figure \ref{fig:gamma} shows the preference of discount factors under different cache sizes, where small cache sizes perform better with relatively smaller $\gamma$ and large cache sizes favor relatively larger $\gamma$.
This is not a coincidence if we dig a little bit deeper into the meaning of $\gamma$.
The discount factor teaches the agent how to balance the immediate reward and the long-term return, i.e., how far the future should be considered when making decisions.
Under caching scenarios in theory, the larger the cache size is, the further future should be taken into account, and thus the relatively larger $\gamma$ is preferred, matching our observations.

Figure \ref{fig:parrot} shows  simple comparison with \textsc{Parrot} on the first 20,000 accesses of 4 MSR traces.
Because \textsc{Parrot} mostly focuses on hardware caches and also relies on $pc$ as one of input features, which does not exist in the software setting, it generally has inferior performance to \phoebe.
For $ts$, \textsc{Parrot} has a higher hit rate than \phoebe;
this is reasonable under some cases, since \textsc{Parrot} is an offline method aware of ample information of traces in advance.
As a completely online RL caching policy, \phoebe\ exhibits its competitiveness by achieving excellent performance close enough to that of \texttt{OPT}.

\section{Related Work}
There are two major venues to apply machine learning (ML) for cache optimization: designing intelligent prefetching policies or improving cache replacement policies.

\textbf{Prefetching.}
To design an intelligent prefetcher, multiple ML techniques can be applied, ranging from the simple perceptron learning \cite{wang2017data,bhatia2019perceptron}, the contextual bandits model in RL \cite{peled2015semantic}, to the more recent LSTM based approaches in which the prefetching problem can be modeled as either a regression problem \cite{zeng2017long} or a classification problem \cite{hashemi2018learning}.
The influence of hyperparameters in LSTM models is also investigated \cite{braun2019understanding} and a neural hierarchical sequence model is developed to accommodate the extremely large memory space \cite{shineural}.

As far as cache in content delivery networks, prefetching is often the focus optimized by different RL techniques, such as combining the Markov decision process (MDP) with side information \cite{somuyiwa2018reinforcement}, multi-armed bandit \cite{blasco2014learning,chang2018learn} and Q-learning \cite{sadeghi2019deep,sadeghi2020reinforcement}.

\textbf{Replacement.}
Most ML based cache replacement policies target on hardware caches, and thus their input features are carefully selected based on hardware characteristics.
In consideration of practical hardware implementation, there is a series of work employing 
perceptron learning \cite{teran2016perceptron,jimenez2017multiperspective}, MDP models \cite{beckmann2017maximizing} and the support vector machine simplified from a complex LSTM model \cite{shi2019applying}.
Though not explicitly mentioned, \textsc{Parrot} \cite{liu2020imitation} also mostly focuses on hardware caches, which is an offline imitation learning based policy with LSTM models.
These efforts, even though displaying great performance, rely heavily on information from hardware characteristics and it is unclear how to extend to software caches.
One state-of-the-art online ML policy is \texttt{LeCaR} \cite{vietri2018driving}, making use of recency and frequency information.

Different from the work reviewed above, \phoebe\ is a reuse-aware RL framework for the optimal \textbf{online} cache replacement, capable to extract adequate information only \textbf{from a single trace} (rather than other hardware based features) and extendable to various cache scenarios. 

\section{Conclusion}
In this work, we propose \phoebe, a reuse-aware RL framework for the optimal online caching via DDPG. 
Through continuous interactions with the cache environment and varying workloads at runtime, \phoebe\ adjusts its policy to approximate the optimal caching policy.
All information required by \phoebe\ can be distilled from a single cache trace, which would make it be universally applicable in theory and can be generalized for different cache scenarios.
In evaluation over a set of Microsoft cloud workloads, \phoebe\ is able to close the gap of cache miss rate from \texttt{LRU} and a state-of-the-art online learning based cache policy to the Belady's optimal policy by 70.3\% and 52.6\%, respectively.

\bibliography{all}

\end{document}